\documentclass[traditabstract]{aa}
\usepackage{natbib}
\bibpunct{(}{)}{;}{a}{}{,} 
\usepackage{txfonts}
\usepackage{graphicx}
\newcommand{\kms}{{\,\rm km\,s}^{-1}} 
 
\newcommand{\nodata}{\multicolumn{1}{c}{   $\cdots$}  }

\renewcommand{\mag}{\mbox{$\;$mag}}

\begin{document}

\title{The luminosity of supernovae of type Ia from TRGB distances and
       the value of \boldmath{$H_{0}$}}

\author{G.~A. Tammann \and B. Reindl}
\institute{%
     Department of Physics and Astronomy, University of Basel,
     Klingelbergstrasse 82, 4056 Basel, Switzerland \\
     \email{g-a.tammann@unibas.ch}}

\date{Received {24 August 2012}}

\abstract{%
     Distances from the tip of the red-giant branch (TRGB) in the halo
     Population of galaxies  -- calibrated through RR~Lyr stars as
     well as tied to Hipparcos parallaxes and further supported by
     stellar models -- are used to determine the luminosity of six
     nearby type Ia supernovae (SN 2011fe, 2007sr, 1998bu, 1989B,
     1972E, and 1937C). The result is 
     $\langle M^\mathrm{corr}_{V}\rangle=-19.41\pm0.05$. If this value
     is applied to 62 SNe\,Ia with $3000< v < 20,000\kms$ a
     large-scale value of the Hubble constant follows of
     $H_{0}=64.0\pm1.6\pm2.0$. The SN \textit{HST} Project gave
     $H_{0}=62.3\pm1.3\pm5.0$ from ten Cepheid-calibrated SNe\,Ia
     (Sandage et~al. 2006). The agreement of young Population~I
     (Cepheids) and old, metal-poor Population~II (TRGB) distance
     indicators is satisfactory. The combined weighted result is 
     $H_{0}=63.7\pm2.3$ (i.e.\ $\pm3.6\%$). The result can also be
     reconciled with WMAP5 data (Reid et~al. 2010).
}
\titlerunning{The luminosity of SNe\,Ia from TRGB distances and $H_{0}$}
\maketitle

\section{Introduction}
\label{sec:1}
The observational determination of the precise large-scale value of
the Hubble constant ($H_{0}$) is decisive for understanding the nature
of dark energy and for determining the equation of state $\omega$ and
other cosmological parameters \citep[e.g.][]{Suyu:etal:12}.
The most direct route to achieve this goal is through the Hubble
diagram of supernovae of type Ia (SNe\,Ia) which are the tightest
standard candles known and trace the linearity of the cosmic expansion
out to $z>1$. But for the numerical value of $H_{0}$ the intrinsic
luminosity of SNe\,Ia is also needed. So far their mean luminosity has 
been derived from a few SNe\,Ia in nearby Population~I parent galaxies
whose distances can be derived from the period-luminosity relation of
Cepheids. Present results from calibrated SNe\,Ia, however, diverge
considerably giving, e.g., $H_{0}=71\pm6$ using only SNe\,Ia from the
\textit{HST} Project for $H_{0}$ \citep{Freedman:etal:01}, 
$62.3\pm5.2$ from the SN \textit{HST} Project \citep{STS:06}, and 
$73.8\pm2.4$ \citep{Riess:etal:11}. 
The divergence is almost entirely caused by different Cepheid distances:
of the ten Cepheid distances used for the calibration of SNe\,Ia by
\citet{STS:06} six are also in the list of \citet{Freedman:etal:01}
and four (out of 8) are in the list of \citet{Riess:etal:11}. The
distances of the two latter samples are shorter on average by 
$0.41\pm0.10$ and $0.32\pm0.08\mag$, respectively, than those by
\citet{Saha:etal:06}. The reason for these differences, that
perpetuate of course into \textit{all} Cepheid-calibrated distance
indicators, are different interpretations of metallicity effects on
the shape and the zero point of the period-color relation and the
period-luminosity relation and the related question of internal
absorption; these points are further discussed by
\citet[][and references therein]{TR:11}. 

     The unsatisfactory situation with Cepheid-calibrated SNe\,Ia
calls for an independent calibration. Therefore an alternative SN\,Ia
calibration is pursued here using tip of the red-giant branch (TRGB)
distances. The two routes to $H_{0}$ are truly independent because
Cepheids belong to the young Population~I whereas the TRGB is a
feature of the old Population~II. 

     Over the last 70 years the TRGB of the \textit{old, metal-poor}
Population has developed into a powerful distance indicator. By now it
fulfills the long-standing hope for a large-scale distance scale based
on only Population~II objects because the inherent disadvantage of the
TRGB, i.e.\ its limited range, can be overcome with the first SNe\,Ia
with known TRGB distances. Thus the TRGB will provide an independent
check on the Population~I distance scale that relies so far almost
entirely on Cepheids.

     The TRGB has several advantages. It is physically well understood
and empirically exceptionally well calibrated. It needs fewer
observations than variable stars, the problem of absorption in the
halo of the host galaxy is minimal, and within the relevant
metallicity range its luminosity is nearly independent of color.
The observation of the TRGB in the galaxian halos obliterate the
danger of blends, a much debated problem in case of Cepheids.  
Moreover the determination of the TRGB -- that is defined as a cutoff
magnitude -- is immune against selection effects (Malmquist bias), a
problem that has severely hampered the application of distance
indicators with substantial internal dispersion. 

     In the following all magnitudes are corrected for Galactic
absorption \citep{Schlegel:etal:98} with $R_{B}=4.1$, $R_{V}=3.1$, and
$R_{I}=1.8$.
Reduced values of 3.65, 2.65, and 1.35, respectively, are adopted for
the absorption of the SNe\,Ia in their host galaxies as further
discussed in Sect.~\ref{sec:3}. The SN magnitudes refer to the
maxima in $B$, $V$, and $I$, respectively, if not otherwise stated.
The colors $(B\!-\!V)^{\max}$ are defined as $(B^{\max} - V^{\max})$.

     The method of the TRGB is discussed in Sect.~\ref{sec:2}.
The TRGB distances and optical luminosities of six SNe\,Ia are derived
in Sect.~\ref{sec:3}.
In Sect.~\ref{sec:4} the calibrated SN\,Ia luminosity is applied to
distant SNe\,Ia leading to the value of $H_{0}$.
$JHK_{s}$ luminosities are briefly discussed in Sect.~\ref{sec:5}.
The discussion and conclusions follow in Sect.~\ref{sec:6}.

\section{The tip of the red-giant branch (TRGB)}
\label{sec:2}
The TRGB was first observed by \citet{Baade:44a,Baade:44b}. He noticed
on red-sensitive plates that in several nearby Population~II galaxies
the resolution into stars suddenly sets in at a fixed luminosity. He
explained what later became known as ``Baade's sheet'' as the top of
the red-giant branch which was then already known in globular clusters. 
The full power of the TRGB became apparent only with the inclusion of
$I$-colors \citep{Mould:Kristian:86,DaCosta:Armandroff:90}. 
The use of the TRGB as an extragalactic distance indicator has been
specifically addressed by \citet{Lee:etal:93,Madore:etal:97}, and many
others.  

     The physical explanation of the TRGB is due to
\citet{Hoyle:Schwarzschild:55} who showed that the maximum luminosity
an old, metal-poor star can reach on the red-giant branch is sharply
bounded -- independent of mass in first approximation -- by the Helium
flash that occurs once the star has developed an electron-degenerate
core. The physics of the TRGB was further elucidated by
\citet{Thomas:67,Rood:72,Sweigart:Gross:78}. 
After the flash low-mass stars settle on the Horizontal Branch and
more massive stars in the Red Clump \citep{Faulkner:66}. 
Later models predict the TRGB luminosity to be nearly constant for
masses below about 1.8 solar masses and $[\textrm{Fe/H}]\!<\!-0.7$.
This holds particularly in the $I$-band where a tip luminosity of
$M^{*}_{I}=-4.05$ is predicted by the models and where the dependence
on the tip color $(B\!-\!V)^{*}$ is minimal
\citep[][and references therein]{Salaris:11}.

\subsection{The empirical calibration of the TRGB}
\label{sec:2:1}
The absolute magnitude of $M^{*}_{I}$ of the TRGB is empirically well
determined for the color range $1.40<(V\!-\!I)^{*}\le1.75$.
\citet{Lee:etal:93} adopted $M^{*}_{I}=-4.0\pm0.1$ from four globular
clusters with RR~Lyr star distances. \citet{Rizzi:etal:07} have fitted
the Horizontal Branches (HB) of five galaxies to the Hipparcos-based
HB of \citet{Carretta:etal:00} allowing for metallicity effects; their
result of $M^{*}_{I}=-4.05\pm0.02$ is independent of RR~Lyr stars. It
is therefore independent of a calibration of the TRGB luminosity from
20 galaxies with known RR~Lyr distances, on the zero point of
\citet{ST:06}, that yields in perfect agreement
$M^{*}_{I}=-4.05\pm0.02$ \citep{TSR:08a}. The individual calibrators
of \citeauthor{Rizzi:etal:07} and \citeauthor{TSR:08a} do not show a
significant color dependence in the indicated color range,
corresponding roughly to a metallicity range of
$-2.0<$[Fe/H]$<-1.1$ in the system of \citet{Carretta:Gratton:97}. 

     The agreement of the TRGB luminosity from model calculations and
observations is so good that the zero-point error is likely to be
$\le0.05\mag$. The dispersion between RR~Lyr and TRGB distances of 20
galaxies is only $0.08\mag$.  
Allowing for a random error of RR~Lyr distances of $0.05\mag$ leaves a
random error for the TRGB distances of only $0.06\mag$.

\subsection{The TRGB in mixed populations}
\label{sec:2:2}
The most severe problem of the TRGB as a distance indicator, as
pointed out already by \citet{Freedman:89}, are very red stars like
supergiants, evolved AGB and carbon stars, that can become brighter
than and as red as the TRGB in dense fields of mixed populations. In
that case the TRGB may get drowned by contaminating stars. This always
leads to an underestimate of the distance. The problem can be
illustrated by the following cases.  

     In case of M101 \citet{Sakai:etal:04} have found, plotting the
red part of the CMD in $I$ versus $(V\!-\!I)$, the clearly visible
TRGB in the \textit{outermost} chips WF3 and WF4 of an outlying WFPC2
field (Fig.\ref{fig:01} left). They applied a Sobel edge-detection kernel
\citep{Lee:etal:93} and adopt $m^{*}_{I}=25.40\pm0.04$ (corrected for
Galactic absorption) for its position. 
The CMD of the entire WFPC2 field, reaching deeper into the disc
population, is swamped with an admixture of evolved AGB stars and
supergiants such that the TRGB becomes invisible to the eye and also
the edge detection spectrum becomes much noisier (Fig.\ref{fig:01} right).
\citet{Rizzi:etal:07} have remeasured the field of
\citet{Sakai:etal:04} and found $m^{*}_{I}=25.29\pm0.08$. The mean of
$m^{*}_{I}=25.35\pm0.05$ is adopted here.
\citet{Shappee:Stanek:11} have suggested $m^{*}_{I}$ to lie at
$24.98\pm0.06$ for a reference color of $(V\!-\!I)^{*}=1.6$, yet the
magnitude is clearly too bright. The value is measured in two
\textit{inner} fields of M101. The authors have taken the precaution
to exclude the stars within $4'75$ from the center, but in spite of
this the field contains a mixture of old and young stars. Clearly some
of the latter are very bright and red and hide the true TRGB. The
authors' edge detection function shows also a weaker maximum at
$m^{*}_{I}\sim25.3$ which presumably is the signature of the true TRGB
in agreement with  \citet{Sakai:etal:04} and \citet{Rizzi:etal:07}.

\def\figsize{0.79}
\begin{figure*}
   \centering
   \resizebox{\figsize\hsize}{!}{\includegraphics{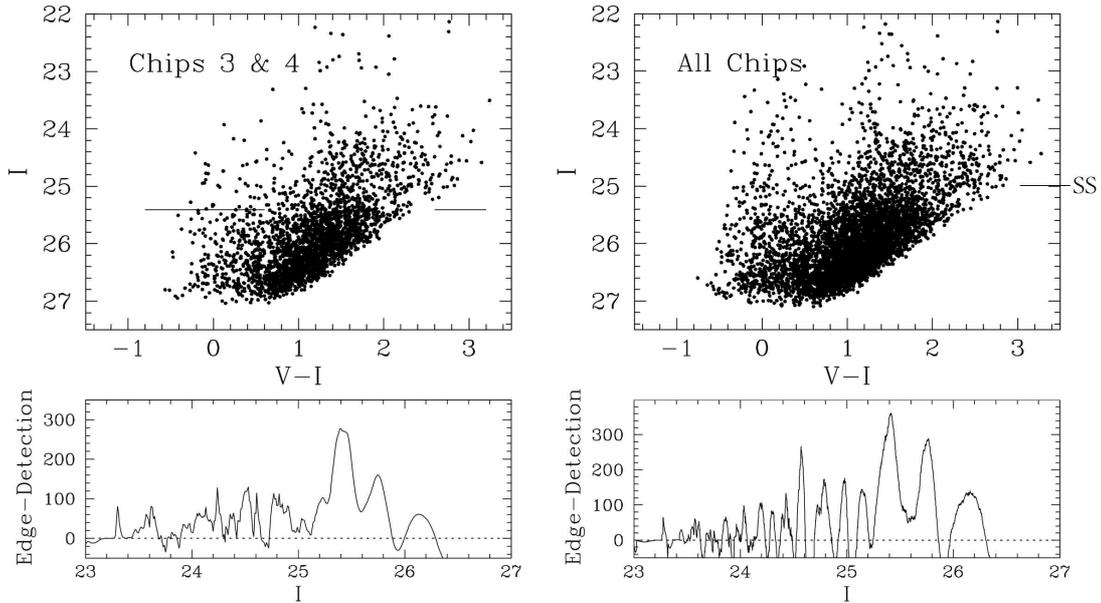}}
   \caption{Color-magnitude diagrams in $I$ versus $(V\!-\!I)$ of an
     outlying WFPC2 field in M101 from \citet[][Fig.11; note that
     the figure legend confuses M101 with NGC\,3621]{Sakai:etal:04}.
     \textit{Left.} 
     Only the stars in the outermost chips WF3 \& WF4 with a
     prevailing halo population are shown. The TRGB is clearly visible
     at $m^{*}_{I}=25.4$. The edge detection spectrum (below)
     confirms the detection with high significance. 
     \textit{Right.}
     All stars in the WFPC2 field are plotted. The additional stars
     have smaller radial distances from M101 and comprise many evolved
     AGB stars and supergiants that populate the region above the
     TRGB. The TRGB is still detectable in the edge detection spectrum
     (below), but with much lower significance. The short horizontal
     bar denoted SS is where \citet{Shappee:Stanek:11} have positioned
     the spurious TRGB for two \textit{inner}, mixed-population fields
     of M101.  
}   
    \label{fig:01}
\end{figure*}

     The situation in NGC\,3621 is similar to M101.
\citet[][Fig.~10]{Sakai:etal:04} have shown that the TRGB is
definitely detected in an outlying WFPC2 field (chip WF2), whereas the
entire WFPC2 field contains so many equally red, but brighter stars
that the TRGB becomes undetectable.

     An additional illustration is provided by the interacting galaxy
NGC\,4038. \citet{Saviane:etal:08}, considering an ACS \textit{HST}
field containing clear tracers of a young population, found in the
quadrant most distant of the star-forming tidal tail two
discontinuities along the red-giant branch, the one at
$m^{*}_{I}=26.65$ and the other one at $\sim\!27.5$. The authors
identified the brighter one with the TRGB. However,
\citet{Schweizer:etal:08} applied a Sobel edge-detection kernel
to the whole ACS field and convincingly showed the TRGB to lie at
$m^{*}_{I}=27.46\pm0.12$; it is the faintest TRGB so far measured. 

     The galaxies NGC\,3368 and NGC\,3627 are bona fide members of the
Leo~I group \citep[e.g.][]{Humason:etal:56,Huchra:Geller:82}, yet
their published $m^{*}_{I}$ magnitudes of 25.61
\citep{Mould:Sakai:09a} and 25.82 \citep{Mould:Sakai:09b},
respectively, are brighter by $\sim\!0.6\mag$ than the mean TRGB
magnitude $\langle m^{*}_{I}\rangle=26.34\pm0.08$ of four unquestioned
group members (NGC\,3351, 3377, 3379, and 3384; see 
\citealt{Sakai:etal:04,Rizzi:etal:07,Mould:Sakai:09a,Mould:Sakai:09b}).
The proposed tip of NGC\,3368 is based on (too) few stars, and the
genuine tip of NGC\,3627 may well be buried in a dense cloud of stars.
The edge detection functions of both galaxies show secondary peaks at
$m^{*}_{I}\sim26.4$ that presumably reflect the true TRGBs. Also the
velocities of $v_{220}=720$ and $428\kms$ ($v_{220}$ is the velocity
corrected for a self-consistent Virgocentric infall model with a local
infall $220\kms$) of NGC\,3368 and NGC\,3627 compared with the
mean velocity of $\langle v_{220}\rangle=497\kms$ ($\sigma=159\kms$)
from 8 bright group members make it unlikely that they could be
foreground galaxies. Moreover, we note that the Cepheid distances of
NGC\,3368 and NGC\,3627 of $(m-M)=30.34$ and $(m-M)=30.50$
\citep{Saha:etal:06} are in very good agreement with the adopted mean
TRGB distance of the Leo~I group of $\langle m-M\rangle=30.39\pm0.08$. 
Finally convincing evidence for the group membership comes from the HI
maps of the Leo~I group \citep[][Figs. 3 and 4]{Stierwalt:etal:09}.
For these arguments NGC\,3368 and 3627 are unquestionable members
of the Leo\,I group; as such they are treated in the following.

     At present about 200 TRGB magnitudes of nearby galaxies have been
determined by many observers \citep[for a compilation see e.g.][in the
following TSR08b]{TSR:08b}. 
First attempts to calibrate SNe\,Ia with TRGBs have already
been made \citep{TSR:08a,Mould:Sakai:09b}.

\def\figsize{0.66}
\begin{figure*}
   \centering
   \resizebox{\figsize\hsize}{!}{\includegraphics{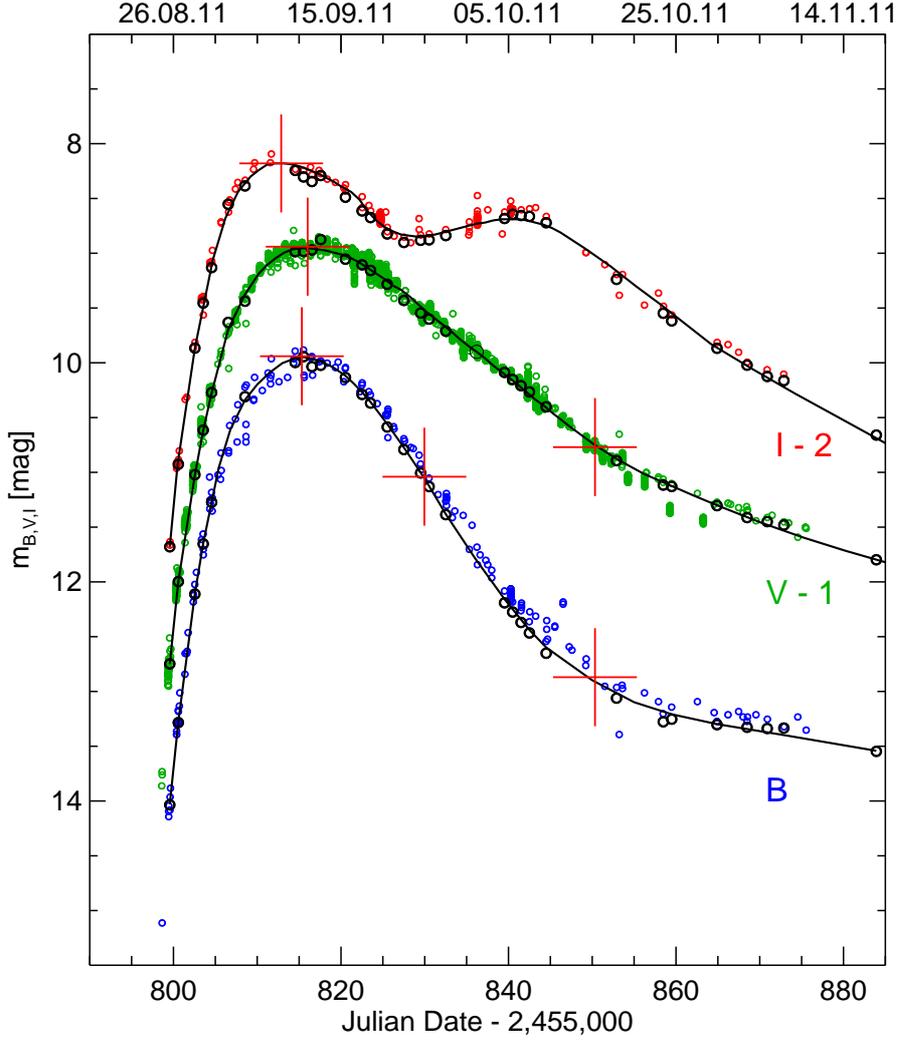}}
   \caption{The light curve of SN\,Ia 2011fe in M101 in $B$ (blue),
     $V$ (green), and $I$ (red) from AAVSO data (small open dots) and
     \citet{Richmond:Smith:12}  (large open dots). The well fitting,
     mean standard light curves in $B$ and $V$ of SNe\,Ia with
     $\Delta m_{15}=1.1$ are overplotted; they are taken from
     \citet{Leibundgut:88}; in $I$ the light curve of SN\,1992al
     \citep{Hamuy:etal:96} is shown for comparison (for details see
     text). The maximum epochs and those 15 and 35 days after $B$ 
     maximum are marked. The light curves in $V$ and $I$ are shifted by
     $-1$ and $-2\mag$, respectively.}   
    \label{fig:02}
\end{figure*}

\section{The six calibrating SNe\,Ia with TRGB distances}
\label{sec:3}
SNe\,Ia develop their full power as standard candles only once they
are fully corrected for absorption and standardized to a fixed decline
rate (or some other light curve characteristics). The calibrating
SNe\,Ia as well as the distant SNe\,Ia, that need in addition
$z$-dependent time dilation and $K$-corrections, are here homogenized
following \citet[][in the following RTS05]{RTS:05} because the
resulting magnitudes define the Hubble diagram with a particularly
small dispersion of $0.14\mag$; even the dispersion of $0.13\mag$ in
the $H$ band is hardly smaller \citep{Kattner:etal:12}.
Other standardization systems have been applied, but the magnitudes
from different systems should not be mixed because they do not
necessarily have the same zero point. In particular, different
assumptions on the intrinsic color of SNe\,Ia and on the
absorption-to-reddening ratio ${R}$ lead to different corrections for
internal absorption and hence to different corrected magnitudes.
However the choice of the intrinsic color and the value of ${R}$
becomes irrelevant as long as nearby calibrators and distant SNe\,Ia
are treated consistently.

     \citeauthor*{RTS:05} have corrected the SN magnitudes in $B$, $V$
and $I$ for Galactic absorption following \citet{Schlegel:etal:98} and
for internal absorption. The \textit{internal} absorption $A_{V}$ is 
$R_{V}\times E(B\!-\!V)$  where $R_{V}=2.65$ is adopted. Several
investigations indicate that such a low value is appropriate for
SNe\,Ia  (e.g. 
\citealt{Branch:Tammann:92}; \citealt{Riess:etal:96};
\citealt{Tripp:98}; \citealt{Krisciunas:etal:00};
\citealt{Altavilla:etal:04}; \citeauthor*{RTS:05};
\citealt{Wang:etal:06}; \citealt{Jha:etal:07}).
The observed colors are defined as $m_{B}^\mathrm{max} -
m_{V}^\mathrm{max}$ and $m_{V}^\mathrm{max} - m_{I}^\mathrm{max}$;
the color $(B\!-\!V)_{35}$, 35 days after $B$ maximum is also used
because the colors \citep{Lira:95,Phillips:etal:99} and spectra
\citep{Filippenko:97} of SNe\,Ia with different decline rates are
quite similar at that epoch \citep[see also][]{Jha:etal:07}. The
reddenings are the differences between the observed colors and the
intrinsic colors. The intrinsic colors $(B\!-\!V)^\mathrm{max}$ ,
$(B\!-\!V)_{35}$, and $(V\!-\!I)$ were derived in function of the 
decline rate $\Delta m_{15}$ from 34 unreddened SNe\,Ia in E/S0
galaxies and from outlying SNe\,Ia in spirals. The resulting excesses
$E(B\!-\!V)_{35}$ and $E(V\!-\!I)$ are converted into values of
$E(B\!-\!V)$ at maximum, using empirical relations
(\citeauthor*{RTS:05}, Eqs. (9) and (12)), and then averaged.
Finally the magnitudes were standardized to a fiducial decline rate of
$\Delta m_{15}=1.1$ (\citeauthor*{RTS:05}, Fig.~12). The magnitudes,
fully corrected for Galactic and host absorption and standardized to
$\Delta m_{15}=1.1$, are denoted $m^\mathrm{corr}$.

\subsection{The SN\,2011fe in M101}
\label{sec:3:1}
The supernova 2011fe was discovered in M101 by the Transient Palomar
Factory on August 24, 2011, and classified as a normal type Ia
\citep{Nugent:etal:11} that had exploded two days before. The American
Association of Variable Star Observers (AAVSO) had collected by 
November 10, 2011, 190 $B$, 4143 $V$, and 252 $I$ magnitudes from 48
different sources. Subsequently published, extensive $UBVRI$ CCD
photometry, starting 16.4 days before $B$ maximum, was provided by
\citet{Richmond:Smith:12} and proves the excellent quality of the
AAVSO data. In addition the Las Cumbres Observatory Global Telescope
Network (LCOGT) will provide daily multi-wavelength photometry of
SN\,2011fe including observations only hours after the explosion by
the Palomar Transient Factory \citep{Bianco:etal:12}; pre-publication
maximum magnitudes and $\Delta m_{15}$ were kindly made available by
F.~Bianco (private communication). The three sources agree very well,
giving $B_\mathrm{max}=9.93\pm0.01$, $V_\mathrm{max}=9.93\pm0.01$,
$I_\mathrm{max}=10.18\pm0.03$, and $\Delta m_{15}=1.10\pm0.06$.
The $BVI$ photometry of \citet{Richmond:Smith:12} and of the AAVSO are
plotted in Fig.~\ref{fig:02} as well as a templet light curve for
$\Delta m_{15}=1.10$ \citep{Leibundgut:88,Hamuy:etal:96} for
comparison. The rising branch of these light curves is drawn here
through the observations of SN\,2011fe itself, and in addition the $V$
light-curve is slightly lowered around the epoch of 40 days after $B$
maximum to better fit the observations.

     The mean color excess from $E(B\!-\!V)$, $E(B\!-\!V)_{35}$ , and
$E(V\!-\!I)$ gives, after appropriate transformation, $\langle
E(B\!-\!V)\rangle=0.01\pm0.02$; this is so close to zero that it is
assumed that the internal absorption of SN\,2011fe is zero. Note that
any absorption correction could make the intrinsic luminosity of
SN\,2011fe only brighter. -- Since SN\,2011fe has closely the fiducial
decline rate of $\Delta m_{15}=1.10$ the correction for the decline
rate is $\delta\Delta m_{15}=0.00\pm0.04$. 

     The mean TRGB magnitude of M101 is well determined in an outer
halo field by \citet{Sakai:etal:04} and \citet{Rizzi:etal:07} to be
$m_{I}^{*}=25.35\pm0.04$ (see Sect.~\ref{sec:2:2}). This with the
above TRGB calibration gives a distance modulus of
$(m-M)=29.40\pm0.06$. With $m_{V}^\mathrm{corr}=9.93\pm0.08$ follows an
absolute magnitude of SN\,2011fe of $M_{V}^\mathrm{corr}=-19.47\pm0.10$. 
The relevant parameters of SN\,2011fe are listed in
Table~\ref{tab:01}.

\begin{figure*}
   \centering
   \resizebox{\figsize\hsize}{!}{\includegraphics{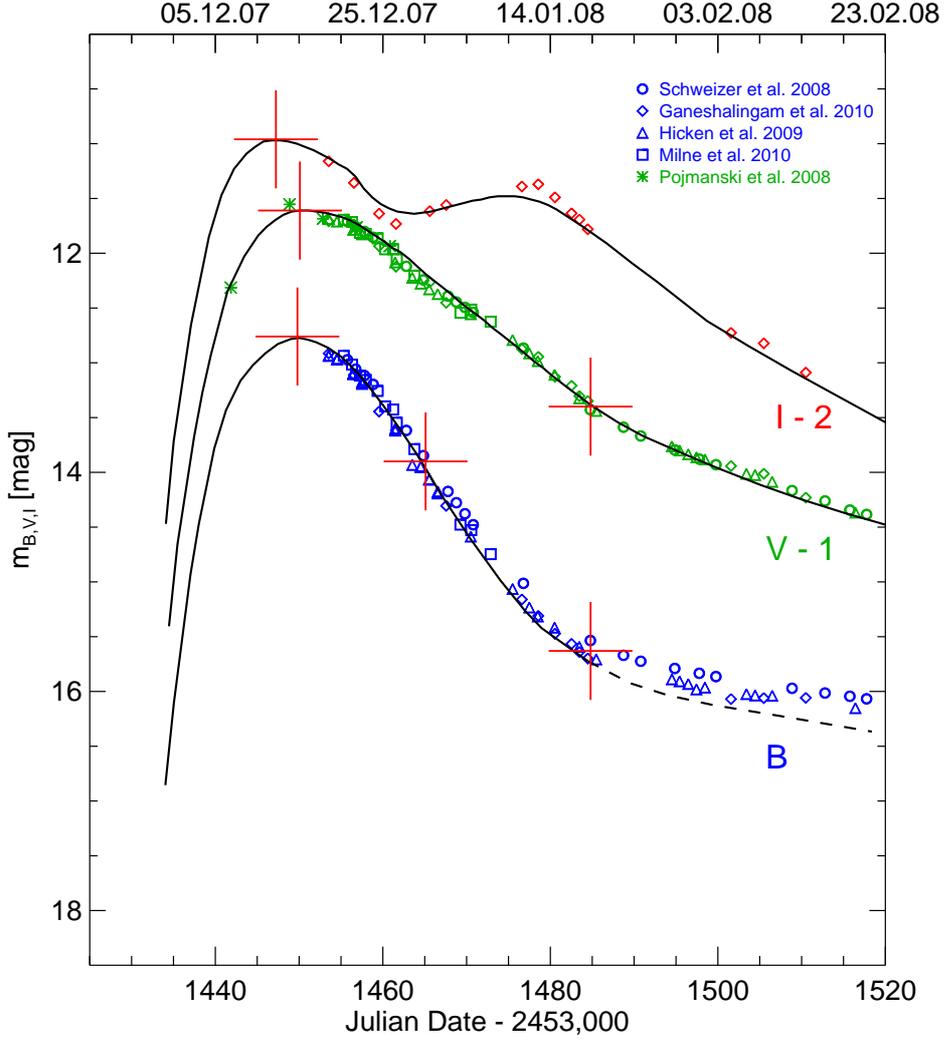}}
   \caption{The light curve of SN\,Ia 2007sr in NGC\,4038 in $B$ (blue),
     $V$ (green), and $I$ (red) from different sources. 
     The full lines are the light curves of SN\,2011fe
     (Fig.~\ref{fig:02}) which has closely the same decline rate of
     $\Delta m_{15}=1.10$; the poorly fitting tail of the $B$ light
     curve is shown as a dashed line. The maximum epochs and those 15
     and 35 days after $B$ maximum are marked. The light curves in $V$
     and $I$ are shifted by $-1$ and $-2\mag$, respectively.}   
    \label{fig:03}
\end{figure*}

\subsection{The SN\,2007sr in NGC\,4038}
\label{sec:3:2}
The supernova 2007sr was discovered as a transient object on the 
southern tail of the interacting galaxy NGC\,4038 on December 12, 2007
\citep{Drake:etal:07}. It was identified as a SN\,Ia by
\citet{Naito:etal:07} and \citet[][see also
\citet{Silverman:etal:12}]{Umbriaco:etal:07}.
The photometry of the object in the $BVI$ bands is here collected; the
data, including the three earliest observations of 
\citet{Pojmanski:etal:08} are plotted in Fig.~\ref{fig:03}, where also
the sources are identified. The data are over-plotted with the
light-curves of SN\,2011fe (Fig.~\ref{fig:02}). They fit the
observations well in $B$ and $V$, except the tail in $B$. The good fit
in $B$ at earlier epochs implies $\Delta m_{15}\!=\!1.10\!\pm\!0.05$. 
The data meander in $I$ about the light curve of SN\,2011fe which is
not unusual for this passband. The adopted $BVI$ maximum magnitudes in
Table~\ref{tab:01} agree within the errors with those of
\citet{Ganeshalingam:etal:10}, that are based on fewer observations.
The maximum value in $B$ is the same as in \citet{Neill:etal:09}. 

     The mean color excess is $\langle E(B\!-\!V)\rangle=0.13\pm0.05$
from $E(B\!-\!V)$ at maximum and $35\;$days after $B$ maximum. The $I$
maximum is too uncertain to be helpful for the determination of the
reddening. The internal absorption becomes then $A_{V}=0.34\pm0.13$
for the adopted value of $R_{V}=2.65$. 

     With the TRGB distance of $(m-M)=31.51\pm0.12$ of
\citet[][see Sect.~\ref{sec:2:2}]{Schweizer:etal:08} 
and with $m_{V}^\mathrm{corr}=12.26\pm0.14$ the absolute magnitude of
SN\,2007sr becomes $M_{V}^\mathrm{corr}=-19.25\pm0.18$. SN\,2007sr is 
the most distant calibrator of the present sample. Its light curve
parameters of interest are compiled in Table~\ref{tab:01}.

\begin{table*}
\begin{center}
\caption{SNe\,Ia with TRGB distances and their luminosities}
\label{tab:01}  
\scriptsize
\begin{tabular}{llrrrccrrrrcc}
\hline
\hline
\noalign{\smallskip}
   \multicolumn{1}{c}{SN}                          &
   \multicolumn{1}{c}{Gal.}                        &
   \multicolumn{1}{c}{$m^{\rm max}_{B}$}           &
   \multicolumn{1}{c}{$m^{\rm max}_{V}$}           &
   \multicolumn{1}{c}{$m^{\rm max}_{I}$}           &
   \multicolumn{1}{c}{$\langle E(B\!-\!V)\rangle$} &
   \multicolumn{1}{c}{$A_{V}$}                     &
   \multicolumn{1}{c}{$\Delta m_{15}$}             &
   \multicolumn{1}{c}{$\delta(\Delta m_{15})$}     &
   \multicolumn{1}{c}{$m^{\rm corr}_{V}$}          &
   \multicolumn{1}{c}{$(m\!-\!M)_{\rm TRGB}$}      &
   \multicolumn{1}{c}{Ref.}                        &
   \multicolumn{1}{c}{$M^{\rm corr}_{V}$}          \\
   \multicolumn{1}{c}{(1)}                         &
   \multicolumn{1}{c}{(2)}                         &
   \multicolumn{1}{c}{(3)}                         &
   \multicolumn{1}{c}{(4)}                         &
   \multicolumn{1}{c}{(5)}                         &
   \multicolumn{1}{c}{(6)}                         &
   \multicolumn{1}{c}{(7)}                         &
   \multicolumn{1}{c}{(8)}                         &
   \multicolumn{1}{c}{(9)}                         &
   \multicolumn{1}{c}{(10)}                        &
   \multicolumn{1}{c}{(11)}                        &
   \multicolumn{1}{c}{(12)}                        &
   \multicolumn{1}{c}{(13)}                        \\
\noalign{\smallskip}
\hline
\noalign{\smallskip}
 2011fe & N5457   & $ 9.93\,(03)$ & $ 9.93\,(03)$ & $10.18\,(04)$ & $0.01\,(02)$ & $0.00\,(05)$ & $1.10\,(06)$ & $ 0.00\,(04)$ & $ 9.93\,(06)$ & $29.40\,(05)$ & 1,2 & $-19.47\,(08)$ \\
 2007sr & N4038   & $12.76\,(05)$ & $12.60\,(05)$ & $12.96\,(08)$ & $0.13\,(05)$ & $0.34\,(13)$ & $1.10\,(05)$ & $ 0.00\,(03)$ & $12.26\,(13)$ & $31.51\,(12)$ &  3  & $-19.25\,(18)$ \\
 1998bu & N3368   & $12.12\,(04)$ & $11.78\,(04)$ & $11.60\,(06)$ & $0.28\,(04)$ & $0.74\,(11)$ & $1.15\,(05)$ & $-0.03\,(03)$ & $11.01\,(12)$ & $30.39\,(10)$ &  4  & $-19.38\,(16)$ \\
 1989B  & N3627   & $12.21\,(07)$ & $11.89\,(05)$ & $11.64\,(06)$ & $0.31\,(03)$ & $0.82\,(08)$ & $1.31\,(05)$ & $-0.13\,(03)$ & $10.94\,(11)$ & $30.39\,(10)$ &  4  & $-19.45\,(15)$ \\
 1972E  & N5253   & $ 8.29\,(10)$ & $ 8.34\,(05)$ & \nodata       & $-0.05$      & $0.00$       & $1.05\,(10)$ & $+0.03\,(06)$ & $ 8.37\,(11)$ & $27.79\,(10)$ & 1,2 & $-19.42\,(15)$ \\
 1937C  & I4182   & $ 8.78\,(15)$ & $ 8.83\,(15)$ & \nodata       & $ 0.00$      & $0.00$       & $0.96\,(10)$ & $+0.09\,(16)$ & $ 8.92\,(16)$ & $28.21\,(05)$ & 1,2 & $-19.29\,(17)$ \\
\noalign{\smallskip}
\hline
\noalign{\smallskip}
\multicolumn{12}{l}{straight mean}             & $-19.38\pm0.04$ \\
\multicolumn{12}{l}{weighted mean}             & $-19.41\pm0.05$ \\
\noalign{\smallskip}
\hline
\end{tabular}
\end{center}
\tablefoot{%
   (1) \citealt{Sakai:etal:04};
   (2) \citealt{Rizzi:etal:07};
   (3) \citealt{Schweizer:etal:08};
   (4) Galaxy assumed at the mean TRGB distance of the Leo~I group
   (see Sect.~\ref{sec:2:2}).
}
\end{table*}

\subsection{The SN\,1998bu in NGC\,3368}
\label{sec:3:3}
The supernova was confirmed to be of type Ia by
\citet{Silverman:etal:12}. The relevant $BVI$ light curve parameters,
based on very extensive photometry by 
\citet{Suntzeff:etal:99,Jha:etal:99,Hernandez:etal:00},
were derived in \citeauthor*{RTS:05} and are listed in
Table~\ref{tab:01}. \citet{Wang:etal:06} and \citet{Hicken:etal:09}
have determined the same ($\pm0.02\mag$)  maxima in $B$ and $V$ from 
an independent templet fitting method. The large reddening is well
determined to be $\langle E(B\!-\!V)\rangle=0.28\pm0.04$  from
$E(B\!-\!V)$ proper, $E(V\!-\!I)$, and $E(B\!-V)_{35}$; this yields
$A_{V}=0.74\pm0.11$. The decline rate $\Delta m_{15}=1.15\pm0.05$ is
close to the standard value of 1.10 and requires a luminosity
correction of only $-0.03\pm0.03\mag$.  

     The resulting value of $m_{V}^\mathrm{corr}=11.01\pm0.12$ gives
-- together with the mean TRGB distance of the Leo~I group, of which
NGC\,3368 is a bona fide member (see Sect.~\ref{sec:2:2}), of
$(m-M)=30.39\pm0.10$ -- an absolute magnitude of
$M_{V}^\mathrm{corr}=-19.38\pm0.16$.

\subsection{The SN\,1989B in NGC\,3627}
\label{sec:3:4}
The supernova was confirmed to be a Branch-normal SN\,Ia by
\citet{Silverman:etal:12}. The $BVI$ light curve parameters are mainly
based on CCD photometry of \citet{Wells:etal:94} and compiled in
\citeauthor*{RTS:05}. The maximum magnitudes agree within $0.02\mag$
with independent templet fits by \citet{Wang:etal:06} and in $B$ and
$V$ also by \citet{Altavilla:etal:04} and \citet{Hicken:etal:09}. The
mean value of $\langle E(B\!-\!V)\rangle=0.31\pm0.03$ from
$E(B\!-\!V)$, $E(V\!-\!I)$ and $E(B\!-\!V)_{35}$ is reliable and
corresponds to $A_{V}=0.82\pm0.08$. The SN was relatively fast
($\Delta m_{15}=1.35$) and requires a correction of the $V$ magnitude
of $-0.13\pm0.03\mag$.

     As shown in Table~\ref{tab:01} the corrected apparent magnitude
becomes $m_{V}^\mathrm{corr}=10.94\pm0.10$. For reasons given in
Sect.~\ref{sec:2:2} NGC\,3627 is assumed to be at the mean TRGB
distance of the Leo~I group of $(m-M)=30.39\pm0.10$. This then gives
an absolute magnitude of $M_{V}^\mathrm{corr}=-19.45\pm0.15$ of
SN\,1989B.

\subsection{The SN\,1972E in NGC\,5253}
\label{sec:3:5}
The supernova is the prototypical Ia SN \citep{Branch:etal:93}. The
well determined photoelectric $BVI$ light curve parameters in
Table~\ref{tab:01} were derived by \citeauthor*{RTS:05} from the
published photometry 
\citep{Lee:etal:72,Cousins:72,Ardeberg:deGroot:73}.
The $I$ photometry of \citet{van:Genderen:75} sets in too late to
determine a reliable value of $I^\mathrm{max}$. The negative mean color
excess $\langle E(B\!-\!V)\rangle$ is based therefore on only
$E(B\!-\!V)$ measured at maximum and 35 days thereafter. Anyhow the 
assumption of zero internal absorption seems secure for this far
outlying SN. The standardization of $\Delta m_{15}=1.05\pm0.05$ to the
fiducial decline rate of $\Delta m_{15}=1.10$ requires a $V$ magnitude
correction of only $+0.03\pm0.06\mag$. This leads to
$m_{V}^\mathrm{corr}=8.37\pm0.11$ which we adopt. -- The templet fitting
of \citet{Wang:etal:06} yields somewhat brighter magnitudes and a
slower decline rate ($B^\mathrm{max}=8.11\pm0.10$,
$V^\mathrm{max}=8.17\pm0.09$, and $\Delta m_{15}=0.87\pm0.10$). The two
effects nearly compensate and decrease $m_{V}^\mathrm{corr}$ by only
$0.03\mag$. 

     The adopted value of $m_{V}^\mathrm{corr}$ and the TRGB distance
of NGC\,5253 of $27.79\pm0.10$ \citep{Sakai:etal:04,Rizzi:etal:07} lead 
to a luminosity of SN\,1972E of $M_{V}^\mathrm{corr}=-19.42\pm0.15$.

\subsection{The SN\,1937C in IC\,4182}
\label{sec:3:6}
The supernova was the subject of \citet{Baade:Zwicky:38} seminal paper 
and was for a long time the prototype SN\,I \citep{Minkowski:40}. It
was later classified as a Branch-normal SN\,Ia \citep{Branch:etal:93}.
The various contemporary photometry is difficult to transform into a
modern system. Two sets of $BV$ light curve parameters have been
derived \citep{Schaefer:94,Schaefer:96,Pierce:Jacoby:95,Jacoby:Pierce:96};
they disagree dependent mainly on the weight given to the photometry
at early and at late phases. Average light curve parameters are
adopted here as shown in Table~\ref{tab:01}; they agree reasonably 
well with those of \citet{Wang:etal:06}. The assigned large errors
comprise either set of parameters. There is no doubt that SN\,1937C
was quite blue at maximum; the absorption in the host galaxy is
therefore negligible. The somewhat slow decline of 
$\Delta m_{15}=0.96\pm0.10$ requires a $V$ magnitude correction of 
$0.09\pm0.06\mag$.
 
     With $m_{V}^\mathrm{corr}=8.92\pm0.16$ and a TRGB distance of
$(m-M)=28.21$ \citep{Sakai:etal:04,Rizzi:etal:07} the absolute
magnitude of SN\,1937C becomes $M_{V}^\mathrm{corr}=-19.29\pm0.17$ .

 \vspace{10pt}

     The parameters of the six SNe\,Ia are collected together with
their TRGB distances and corresponding absolute magnitudes in
Table~\ref{tab:01}. 
Columns~3$-$5 give the apparent $BVI$ magnitudes corrected for
Galactic absorption. The color excesses in Col.~6 are the mean of
$E(B\!-\!V)$, $E(B\!-\!V)_{35}$, and $E(V\!-\!I)$, the latter two
reduced to the value of $E(B\!-\!V)$ at maximum. The absorption 
$A_{V}$ in Col.~7 is based on $R_{V}=2.65$. Columns~8 and 9 give the
decline rate $\Delta m_{15}$ and the correction 
$\delta(\Delta m_{15})$ for the standardization to 
$\Delta m_{15}=1.10$. The values $m_{V}^\mathrm{corr}$ in Col.~10 are 
the apparent $V$ magnitudes corrected for internal absorption and for
$\delta(\Delta m_{15}$). The adopted TRGB moduli and their sources are
in Cols.~11 and 12. Column~13 lists the final absolute $V$ magnitudes
$M_{V}^\mathrm{corr}$. The statistical errors of the various entries
are shown in parentheses in units of $0.01\mag$.

     The mean luminosity of the six SNe\,Ia is
$M_{V}^\mathrm{corr}=-19.38\pm0.04$. The dispersion of
$\sigma=0.09\mag$ is even smaller than that of the distant SNe\,Ia
($0.14\mag$). The small dispersion suggests that the random errors
assigned to the individual SNe\,Ia are rather overestimated. The
weighted mean luminosity is $M_{V}^\mathrm{corr}=-19.41\pm0.05$ which
we adopt.

\begin{table*}
\begin{center}
\caption{$JHK_{s}$ luminosities of three SNe\,Ia with TRGB distances}
\label{tab:02}  
\footnotesize
\begin{tabular}{lrrrccccc}
\hline
\hline
\noalign{\smallskip}
   \multicolumn{1}{c}{SN}                          &
   \multicolumn{1}{c}{$J$}                         &
   \multicolumn{1}{c}{$H$}                         &
   \multicolumn{1}{c}{$K_{s}$}                     &
   \multicolumn{1}{c}{Ref.}                        &
   \multicolumn{1}{c}{$(m\!-\!M)_{\rm TRGB}$}      &
   \multicolumn{1}{c}{$M_{J}$}                     &
   \multicolumn{1}{c}{$M_{H}$}                     &
   \multicolumn{1}{c}{$M_{K_{s}}$}                 \\
   \multicolumn{1}{c}{(1)}                         &
   \multicolumn{1}{c}{(2)}                         &
   \multicolumn{1}{c}{(3)}                         &
   \multicolumn{1}{c}{(4)}                         &
   \multicolumn{1}{c}{(5)}                         &
   \multicolumn{1}{c}{(6)}                         &
   \multicolumn{1}{c}{(7)}                         &
   \multicolumn{1}{c}{(8)}                         &
   \multicolumn{1}{c}{(9)}                         \\
\noalign{\smallskip}
\hline
\noalign{\smallskip}
 2011fe & $10.51\,(04)$ & $10.75\,(04)$ & $10.64\,(05)$ & 1 & $29.40\,(05)$ & $-18.89\,(06)$ & $-18.65\,(06)$ & $-18.76\,(07)$ \\
 2007sr & $13.31\,(06)$ & $13.47\,(02)$ & \nodata       & 2 & $31.51\,(12)$ & $-18.20\,(13)$ & $-18.04\,(12)$ & \nodata        \\
 1998bu & $11.55\,(03)$ & $11.59\,(03)$ & $11.42\,(03)$ & 3 & $30.39\,(10)$ & $-18.84\,(10)$ & $-18.80\,(10)$ & $-18.97\,(10)$ \\
\noalign{\smallskip}
\hline
\end{tabular}
\end{center}
\tablefoot{%
   (1) \citealt{Matheson:etal:12};
   (2) \citealt{Schweizer:etal:08};
   (3) \citealt{Wood-Vasey:etal:08}.
}
\end{table*}

\section{The calibration of \boldmath{$H_{0}$}}
\label{sec:4}
The $\log cz$ -- apparent-magnitude relations of a $\Lambda$CDM model
for the case $\Omega_\mathrm{M}=0.3$ and $\Omega_{\Lambda}=0.7$ is given
here \citep{Carroll:etal:92}

\begin{equation}\label{eq:hubble:01}
   \phi(z) = 0.2 m_{\lambda}^\mathrm{corr} + C_{\lambda},
\end{equation}
where
\begin{equation}\label{eq:hubble:02}
   \phi(z) = \log c(1+z_{1})\int_{0}^{z_{1}}[(1+z)^{2}(1+0.3z)-0.7z(2+z)]^{-1/2}dz,
\end{equation}
and where the intercept is given by
\begin{equation}\label{eq:hubble:03}
   C_{\lambda} = \log H_{0} - 0.2 M_{\lambda}^\mathrm{corr} -5,
\end{equation}
from which follows
\begin{equation}
     \log H_{0} = 0.2M^\mathrm{corr}_{V} + C_{V} + 5.
\label{eq:04}
\end{equation}
The intercept $C_{V}$ of the Hubble line was determined by 62
standardized SNe\,Ia with $3000<v<20,000\kms$ to be
$C_{V}=0.688\pm0.004$ (\citeauthor*{RTS:05}, Table~9). Inserting this
value and the weighted mean absolute magnitude $M^\mathrm{corr}_{V}
=-19.41\pm0.05$ of the six SNe\,Ia from Table~\ref{tab:01} gives  
\begin{equation}
        H_{0} = 64.0\pm1.6\pm2.0.
\label{eq:05}
\end{equation}
The random error is small (2.5\%). Systematic errors are due to errors
of the photometry ($\pm0.05\mag$), to the zero-point error of the
adopted TRGB ($\pm0.05$), to the error of the corrections for internal
absorption, and to errors of the standardization to $\Delta m_{15}$.
For the \textit{internal} absorption the value of $R_{V}$ was assumed
to be 2.65 throughout. The average reddening of the calibrating
SNe\,Ia is $E(B\!-\!V)=0.12$ and of the distant SNe\,Ia defining the
Hubble line $E(B\!-\!V)=0.06$. Only the difference of these average
reddenings multiplied with the error of $R$ enter the systematic
error. Even for a wide range of $2.1<R<3.2$ the systematic error is
therefore only $0.04\mag$. The mean $\Delta m_{15}$ values of the six
SNe\,Ia in Table~\ref{tab:01} and of the 62 distant SNe\,Ia are
$1.11\pm0.05$ and $1.22\pm0.03$, respectively. The difference of
$0.11\pm0.06$ multiplied with the error of the slope of the Deltam15
correction ($0.07$) enters as a systematic error which in this case is
confined to $0.01\mag$. The compounded systematic error is therefore
$0.064\mag$ (3.2\%). 

     The solution in equation~(\ref{eq:05}) is stable against the
selection of the SNe\,Ia in Table~\ref{tab:01}. The exclusion of any
two or any three SNe\,Ia yields mean values of $H_{0}$ between 62.8
and 67.3. The three SNe\,Ia with relatively large internal-absorption
corrections and those with zero internal-absorption corrections yield
the same mean value of $H_{0}$ within 2\%. Even under the devious
assumption of zero internal absorption for all six SNe\,Ia their mean
luminosity decreases by only $0.11\mag$, increasing $H_{0}$ by 6\%.

     Moreover, the effect of the adopted cosmological model on $H_{0}$
is small because the distant SN sample is confined to
$\le20,000\kms$. Replacing the $\Lambda$CDM model used here with a
Euclidean model gives an intercept of the Hubble line of $C_{V}=0.679$
and decreases $H_{0}$ by only 1.3 units.

     Summarizing the random and systematic errors added in quadrature
gives a safe upper limit for the total error of $H_{0}$ of 5\%.

\section{A note on the \boldmath{$JHK_{s}$} luminosities of SNe\,Ia}
\label{sec:5}
For three of the SNe\,Ia in Table~\ref{tab:01} also near-infrared
maximum magnitudes in $J$ and $H$ are available, for SN\,2011fe and
SN\,1998bu also in $K_{s}$. The data are compiled in
Table~\ref{tab:02} where also the TRGB distances are repeated from
Table~\ref{tab:01} and the ensuing absolute magnitudes are 
given. The absolute magnitudes of the two last-mentioned SNe\,Ia agree
in $JHK_{s}$ to within $0.05\pm0.12$, $0.15\pm0.12$, and $0.21\pm0.12$,
respectively. Yet any correction for internal absorption, one wants to
apply, can only increase these differences, because SN\,2011fe is
nearly absorption-free whereas SN\,1998bu has a color excess of
$E(B\!-\!V)=0.28$ (Sect.~\ref{sec:3:3}). 


     \citet{Folatelli:etal:10} have calculated the mean absolute
maximum $JHK_{s}$ magnitudes of a set of SNe\,Ia from their velocity
distances assuming $H_{0}=72$ and applying no correction for internal
absorption. If their result is scaled to the above value of
$H_{0}=64.0$ one obtains $M_{J}=-18.69\pm0.07$, $M_{H}=-18.68\pm0.06$, 
and $M_{K_{s}}=-18.73\pm0.12$. This is slightly fainter (1$-$2$\sigma$)
than the mean magnitude of SN\,2011fe and SN\,2007sr
($M_{J}=-18.87\pm0.06$, $M_{H}=-18.73\pm0.06$, and
$M_{K_{s}}=-18.87\pm0.07$) and would suggest that $H_{0}$ should be
further reduced by a modest amount. But it would be imprudent to
attempt an infrared luminosity calibration of SNe\,Ia with only two
objects. 
        
     As seen in Table~\ref{tab:02} SN\,2007sr appears to be less
luminous in $J$ and $H$ by as much as $\sim\!0.7\mag$ on the basis of
the published maximum magnitudes \citep{Schweizer:etal:08} than the
two preceding SNe. This is surprising since it is spectroscopically a
normal SNIa with $B,V$ lightcurves quite similar to SN\,2011fe and
other SNe\,Ia. It can hardly be the effect of internal absorption
because its optical color excess is intermediate ($E(B\!-\!V)=0.13$;
Sect.~\ref{sec:3:2}). However the $J$ and $H$ observations set in only
5 days after $B$ maximum and hence roughly 10 days after the expected
infrared maxima; for the true $J$ and $H$ maxima remains therefore a
considerable margin.

\section{Discussion}
\label{sec:6}
SNe\,Ia play a special role for the determination of the cosmic value
of $H_{0}$ because no other (relative) distance indicators trace the
cosmic expansion field with so small dispersion well beyond all
possible local effect. The next best route is offered by galaxy
clusters with multiple $I$-band Tully-Fisher distances, although this
method has large intrinsic dispersion and is therefore particularly
susceptible to selection bias. The group of R. Giovanelli has
determined such distances for 28 clusters with 26 cluster members per
cluster on average \citep{Masters:etal:06}. After a very careful and
complex correction for various selection effects they define the slope
of the galaxy-type-dependent $m_{I}-\log(\rm{line width})$ relation.
The ensuing \textit{relative} cluster distances define the Hubble line
with a surprisingly small scatter of only $0.15\mag$ over a velocity
range from 2000 to just $10,000\kms$. However the three nearest
clusters with $v<2000\kms$ lie $0.4\mag$ below the Hubble line
\citep[][Fig.~8]{TSR:08a} implying a 20\% decrease of $H_{0}$ around 
$2000\kms$, which is impossible in view of the linearity of the Hubble
flow \citep[e.g.][]{SRT:10}. The discontinuity of $H_{0}$ is a clear
indication that the objective definition of cluster samples could not
be maintained in nearby clusters. \citet{Masters:etal:06} have
zero-pointed their relative cluster distances by means of 15 nearby
field galaxies with Cepheid distances from \citet{Freedman:etal:01}
and of two additional galaxies from other sources. (Their Cepheid
moduli carry a special metallicity correction and are larger on
average than those given by \citet{Freedman:etal:01} by $0.10\mag$;
they are still smaller than those given by \citet{Saha:etal:06} by
$0.15\mag$ on average). In this way they derived a value of
$H_{0}=74\pm2$. However, we strongly object to the calibration
procedure: the objectively homogenized cluster sample cannot be
compared with the sample of nearby calibrators that are chosen for the
sole reason that Cepheid distances and suitable line width data are
available for them. 

     \citet*{Hislop:Mould:11} have used the data for 15 clusters from
\citet{Masters:etal:06}, but their $I$-band Tully-Fisher relation is
calibrated with 13 local field galaxies with available TRGB distances.
Six of these calibrators (NGC 247, 3351, 3368, 3621, 3627, and 4826)
are questionable because the respective TRGB distances published by
\citet{Mould:Sakai:08,Mould:Sakai:09a,Mould:Sakai:09b} and
\citet{Sakai:etal:04} are larger by $0.34\mag$ on average.
But this is of minor importance compared with the principal objection
that -- as in the case of \citet{Masters:etal:06} -- the samples of
calibrators and cluster galaxies constitute entirely differently
selected samples. If anything, the situation is here even worse, because
\citeauthor{Hislop:Mould:11} use only ~30\% of the available cluster
members thereby questioning the corrections for selection bias
derived by \citet{Masters:etal:06}. For these reasons the value of
$H_{0}=79\pm2$ of \citet{Hislop:Mould:11}  should be given low weight.

     To support the present solution it is noted that for the six
galaxies in Table~\ref{tab:01} also Cepheid distances are available.
The corresponding data are set out in the self-explanatory
Table~\ref{tab:03}. The resulting mean Cepheid-calibrated luminosity
of the six SNe\,Ia is $M_{V}^\mathrm{corr}=-19.40\pm0.06$ with a
dispersion of $0.15\mag$ (i.e.\ larger than in case of the TRGB
calibration). Hence the calibration through the TRGB distances and the
adopted Cepheid distances yields -- in fortuitous agreement -- the
same mean luminosity of SNe\,Ia to within $0.01\mag$.

\begin{table}
\begin{center}
\caption{The luminosity of six calibrating SNe\,Ia from Cepheid distances}
\label{tab:03}  
\scriptsize
\begin{tabular}{llrccc}
\hline
\hline
\noalign{\smallskip}
   \multicolumn{1}{c}{SN}                          &
   \multicolumn{1}{c}{Gal.}                        &
   \multicolumn{1}{c}{$m^{\rm corr}_{V}$}          &
   \multicolumn{1}{c}{$(m\!-\!M)_{\rm Ceph}$}      &
   \multicolumn{1}{c}{Ref.}                        &
   \multicolumn{1}{c}{$M^{\rm corr}_{V}$}          \\
   \multicolumn{1}{c}{(1)}                         &
   \multicolumn{1}{c}{(2)}                         &
   \multicolumn{1}{c}{(3)}                         &
   \multicolumn{1}{c}{(4)}                         &
   \multicolumn{1}{c}{(5)}                         &
   \multicolumn{1}{c}{(6)}                         \\
\noalign{\smallskip}
\hline
\noalign{\smallskip}
 2011fe & N5457   & $ 9.93\,(09)$ & $29.28\,(08)$ & 1 & $-19.35\,(11)$ \\
 2007sr & N4038   & $12.28\,(17)$ & $31.66\,(08)$ & 2 & $-19.38\,(19)$ \\
 1998bu & N3368   & $11.01\,(12)$ & $30.34\,(11)$ & 3 & $-19.33\,(16)$ \\
 1989B  & N3627   & $10.94\,(11)$ & $30.50\,(09)$ & 3 & $-19.56\,(14)$ \\
 1972E  & N5253   & $ 8.37\,(11)$ & $28.05\,(27)$ & 3 & $-19.68\,(29)$ \\
 1937C  & I4182   & $ 8.92\,(16)$ & $28.21\,(09)$ & 3 & $-19.29\,(18)$ \\
\noalign{\smallskip}
\hline
\noalign{\smallskip}
\multicolumn{5}{l}{straight mean}             & $-19.43\pm0.06$ \\
\multicolumn{5}{l}{weighted mean}             & $-19.40\pm0.06$ \\
\noalign{\smallskip}
\hline
\end{tabular}
\end{center}
\tablefoot{%
   (1) \citealt{TR:11};
   (2) \citealt{Riess:etal:11} from $H$ magnitudes;
   (3) \citealt{Saha:etal:06}.
}
\end{table}

     It is also noted that the TRGB distances of 78 field
galaxies with $v_{220}>280\kms$ give a local value of $H_{0}=62.9\pm1.6$ 
(\citeauthor*{TSR:08b}). A slightly augmented sample of
\citet{Saha:etal:06}, comprising now 29 Cepheid distances with
$280<v_{220}<1600\kms$, yields $63.4\pm1.8$ (\citeauthor*{TSR:08b}).
The large-scale value from the summary paper of the \textit{HST}
Supernova Project, based on ten SNe\,Ia with Cepheid distances and a
total of 62 standardized SNe\,Ia with $2000<v<20,000\kms$, that
resulted in $H_{0}=62.3\pm1.3\pm5.0$ \citep{STS:06}, has been
mentioned before. 

     Additional support for a rather low value of $H_{0}$ comes from
\citet{Reid:etal:10} who have derived -- on the assumption of a
$\Lambda$CDM model, without using $H_{0}$ as a prior -- a value of
$H_{0}=65.6\pm2.5$ by combining the 5-year WMAP data
\citep{Dunkley:etal:09} with the Hubble diagram of the SN\,Ia Union
Sample \citep{Kowalski:etal:08} and with a sample of over 110,000
luminous red galaxies (LRG) from the Sloan Digital Sky Survey DR7. 

     A new prospect to \textit{directly} measure $H_{0}$ at
cosmological relevant distances comes from water maser distances of
galaxies beyond $\sim\!3000\kms$; the first two galaxies, UGC\,3789
and NGC\,6264 yield a  mean value of $H_{0}=67$ with a still large
error of $\pm6$ \citep{Braatz:etal:10,Kuo:11,Henkel:etal:12}.
Further progress along this novel route is to be expected.
-- Present results from 
strongly lensed quasars \citep[e.g.][]{Vuissoz:etal:08} and from the 
Sunyaev-Zeldovich effect \citep[][]{Holanda:etal:12} give even larger 
error margins on $H_{0}$.

     The main conclusion is that the TRGB distances of six SNe\,Ia
require a large-scale value of the Hubble constant of
$H_{0}=64.0\pm2.6$ (including random and systematic errors). Suggested
values as high as $H_{0}>70$ are unlikely in the light of the TRGB. If
the present result is combined with the independent value of $H_{0}=
62.3\pm5.2$ from ten Cepheid-calibrated SNe\,Ia \citep{STS:06} 
one obtains a weighted mean of $H_{0}=63.7\pm2.3$, which is our best
estimate. The result is in good agreement with local determinations of
$H_{0}$ and the linearity of the expansion field \citep{SRT:10}. 

     The present result will be further improved by future TRGB
distances of galaxies that have produced a standard SN\,Ia. It is
desirable that the search for the TRGB in NGC\,3368 and 3627 in the
Leo~I group will be repeated in some uncontaminated halo fields in
order to definitely proof their group membership. The next easiest
targets at present are the Branch normal SN\,Ia in the Virgo cluster
galaxies NGC\,4419, 4501 (with the highly absorbed SN\,1999cl), 4526,
and 4639 (yet on the far side of Virgo); the two rather old SNe\,Ia in
NGC\,4496A and 4536 in the W-cloud are less attractive. Somewhat more
difficult are the Fornax cluster galaxies NGC\,1316 with five SNe\,Ia
of which three are Branch-normal 
\citep[1980N, 1981D, and 2006dd;][]{Stritzinger:etal:10}, and the
galaxies NGC\,1380 and NGC\,1448 with one SN\,Ia each; SN\,2007on in
NGC\,1404 is as a fast decliner \citep{Kattner:etal:12} unreliable for
calibration purposes. The Virgo and Fornax galaxies will probably be
easier and more difficult, respectively, by a few $0.1\mag$ than
NGC\,4038 where \citet{Schweizer:etal:08} have demonstrated that the
TRGB can be reached.

     It is foreseeable that the route to a precision value of $H_{0}$
through TRGB-calibrated SNe\,Ia will become highly competitive with any
Cepheid-based distance scale.

\begin{acknowledgements}
We thank Dr. Shoko Sakai for valuable information and Dr. Federica
Bianco for pre-publication data. We have made extensive use of the
excellent data for SN\,2011fe of the AAVSO. 
\end{acknowledgements}



\end{document}